\documentclass[aps,twocolumn,showpacs]{revtex4}
\usepackage{graphicx,bbm,amssymb}
\begin{document}
\title{Experimental reconstruction of photon statistics without photon
counting}
\author{Guido Zambra, Alessandra Andreoni, Maria Bondani}
\email{maria.bondani@uninsubria.it}
\affiliation{Istituto Nazionale per la Fisica della Materia, INFM and
Dipartimento di Fisica e Matematica, Universit\`a degli Studi
dell'Insubria, Como, Italia.}
\author{Marco Gramegna, Marco
Genovese} \email{genovese@ien.it}\author{Giorgio
Brida}\affiliation{Istituto Elettrotecnico Nazionale, IEN, Galileo
Ferraris, Torino, Italia.}\author{Andrea Rossi and Matteo G. A.
Paris} \email{matteo.paris@fisica.unimi.it}
\affiliation{Dipartimento di Fisica dell' Universit\`a di Milano, Italia.}
%%%%%%%%%%%%%%
\begin{abstract}
Experimental reconstructions of photon number distributions of both
continuous-wave and pulsed light beams are reported. Our scheme is
based on on/off avalanche photodetection assisted by
maximum-likelihood estimation and does not involve photon counting.
Reconstructions of the distribution for both semiclassical and
quantum states of light are reported for single-mode as well as for
multimode beams.
\end{abstract}
%%%%%%%%%%%%%%
\maketitle
%%%%%%%%%%%%%%
The measurement of the statistical distribution of the number of
photons provides fundamental information on the nature of any
optical field. The choice of a detector with internal gain suitable
for the measurement is not trivial when the flux of the photons to
be counted is such that more than one photon is detected in the
time-window of the measurement, which is set by the detector
pulse-response, or by an electronic gate on the detector output, or
by the duration of the light pulse. In this case, we need a
congruous linearity in the internal current amplification process:
each of the single electrons produced by the different photons in
the primary step of the detection process (either ionization or
promotion to a conduction band) must experience the same average
gain and this gain must have sufficiently low spread. The
fulfillment of both requisites is necessary for the charge integral
of the output current pulse be proportional to the number of
detected photons. Photon detectors that can operate as photon
counters are rather rare. Among these, PhotoMultiplier Tubes (PMT's)
\cite{burle} and hybrid photodetectors \cite{NIST} have the drawback
of a low quantum efficiency, since the detection starts with the
emission of an electron from the photocathode. Solid state detectors
with internal gain, in which the nature of the primary detection
process ensures higher efficiency, are still under development.
Highly efficient thermal detectors have also been used as photon
counters, though their operating conditions are still extreme
(cryogenic conditions) to allow common use \cite{xxx, serg}. The
advent of quantum tomography provided an alternative method to
measure photon number distributions \cite{mun}. However, the
tomography of a state, which has been applied to several quantum
states \cite{raymerLNP}, needs the implementation of homodyne
detection, which in turn requires the appropriate mode matching of
the signal with a suitable local oscillator at a beam splitter. Such
mode matching is a particularly challenging task in the case of
pulsed optical fields.
\par
Photodetectors that are usually employed in quantum optics such as
Avalanche PhotoDiodes (APD's) operating in the Geiger mode
\cite{rev, serg} seem to be by definition useless as photon
counters. They are the solid state photodetectors with the highest
quantum efficiency but they have the obvious drawback that the
breakdown current is independent of the number of detected photons,
which in turn cannot be determined. The outcome of these APD's is
either "off" (no photons detected) or "on" {\em i.e.} a "click",
indicating the detection of one or more photons. Actually, such an
outcome can be provided by any photodetector (PMT, hybrid
photodetector, cryogenic thermal detector) for which the charge
contained in dark pulses is definitely below that of the output
current pulses corresponding to the detection of at least one
photon. Note that for most high-gain PMT's the anodic pulses
corresponding to no photons detected can be easily discriminated by
a threshold from those corresponding to the detection of one or more
photons.
\par
The statistics of the "no-click" and "click" events
from an on/off detector, assuming no dark counts,
is given by
\begin{eqnarray}
p_0(\eta) &=& \sum_n (1-\eta)^n \varrho_n\: \label{p0}\:,
\end{eqnarray}
and $p_{>0} (\eta) =1- p_0(\eta)$, where $\varrho_n$ is the probability
of finding $n$ photons and $\eta$ is the quantum efficiency of the detector, {\em
i.e.} the probability of a single photon to be revealed.  At first
sight the statistics of an on/off detector appears to provide
quite a scarce piece of information about the state under
investigation. However, if the statistics about $p_0(\eta)$ is
collected for a suitably large set of efficiency values then the
information is enough to reconstruct the whole photon distribution
$\varrho_n$ of the signal, upon a suitable truncation of the Hilbert
space.
\par
The reconstruction of photon distribution through on/off detection
at different efficiencies has been analyzed \cite{mogy} and its
statistical reliability investigated in some details \cite{pcount}.
In addition, the case of few and small values of $\eta$ \cite{ar}
has been addressed. However, whilst these theoretical studies found
an application to realize a multichannel fiber loop detector
\cite{olom,kb}, an experimental implementation of this technique for
reconstructing photon distribution of a free-propagating field is
still missing. In view of the relevance of photon distribution for
applications in quantum information and foundations of quantum
mechanics, the purpose of this letter is to show that a
reconstruction of the photon distribution by using this technique
can be effectively realized experimentally with excellent results
both for free-propagating continuous-wave (cw) and pulsed light
beams, for both single-mode semiclassical and quantum states, as
well as for multimode states.
\par
The procedure consists in measuring a given signal by on/off
detection using different values $\eta_\nu$ ($\nu=1,...,K$) of the
quantum efficiency.  The information provided by experimental data
is contained in the collection of frequencies $f_{\nu} =
f_0(\eta_\nu) = n_{0\nu}/n_\nu$ where $n_{0\nu}$ is the number of
"no click" events and $n_\nu$ the total number of runs with quantum
efficiency $\eta_\nu$.  Then we consider expression (\ref{p0}) as a
statistical model for the parameters ${\varrho_n}$ to be solved by
maximum-likelihood (ML) estimation. Upon defining $p_\nu\equiv
p_0(\eta_\nu)$ and $A_{\nu n} = (1-\eta_{\nu})^n$ we rewrite
expression (\ref{p0}) as $p_{\nu} = \sum_{n} A_{\nu n} \varrho_n$.
Since the model is linear and the parameters to be estimates are
positive (LINPOS problem), then the solution can be obtained by
using the Expectation-Maximization algorithm (EM) \cite{EMalg}. By
imposing the restriction $\sum_n \varrho_n = 1$, we obtain the
iterative solution
\begin{equation}
\varrho_n^{(i+1)} = \varrho_n^{(i)}\sum_{\nu=1}^K \frac{A_{\nu
n}}{\sum_\lambda A_{\lambda n}}
\frac{f_{\nu}}{p_{\nu}[\{\varrho_n^{(i)}\}]}\: \label{ems}\:
\end{equation}
where $p_{\nu}[\{\varrho_n^{(i)}\}]$ are the probabilities
$p_{\nu}$, as calculated by using the reconstructed distribution
$\{\varrho_n^{(i)}\}$ at the $i$-th iteration. As a measure of
convergence we use the total absolute error at the $i$-th iteration
$\varepsilon^{(i)} =\sum_{\nu=0}^K \left| f_\nu- p_\nu [\{
\varrho_n^{(i)}\}]\right|$ and stop the algorithm as soon as
$\varepsilon^{(i)}$ goes below a given level. The total error
measures the distance of the probabilities $p_\nu [\{
\varrho_n^{(i)}\}]$, as calculated at the $i$-th iteration, from the
actual experimental frequencies. As a measure of accuracy we adopt
the fidelity $G^{(i)} = \sum_n \sqrt{\varrho_n \: \varrho_n^{(i)}}$
between the reconstructed distribution and the theoretical one.
\par
In order to verify the potentialities of this technique we applied
it to the reconstruction of various quantum optical states,
generated either in the cw or in the pulsed regimes.
%%%%%%%%%%%%%%%%%%
\begin{figure}[h]
\includegraphics[width=0.3\textwidth]{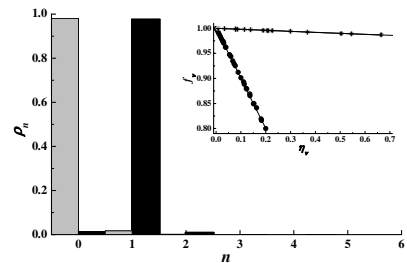}
\caption{\label{f:fock} Reconstruction of the photon distribution
for a weak coherent state (grey) and for the heralded 
single-photon state produced in type II PDC (black). 
Inset: Experimental frequencies $f_{\nu}$
of no-click events as a function
of the quantum efficiency $\eta_{\nu}$ for a weak coherent state
(upper curve) and for PDC heralded photon state (lower curve)
compared with the theoretical curves, $p_\nu \simeq
1-\eta_\nu |\alpha|^2$ and $p_\nu = 1-\eta_\nu$ respectively.}
\end{figure}
%%%%%%%%%%%%%%%%%%
As a first example we have considered single photon states that have
been generated by producing Parametric Down Conversion (PDC)
heralded photons. In little more detail, a pair of correlated
photons has been generated by pumping a type II
$\beta$-Barium-Borate (BBO) crystal with a cw Argon ion laser beam
(351 nm) in collinear geometry. After having split the photons of
the pair by means of a polarizing beam splitter, the detection of
one of the two by a silicon avalanche photodiode detector
(SPCM-AQR-15, Perkin Elmer) was used as an indication of the
presence of the second photon in the other channel, namely a window
of 4.9 ns was opened for detection in arm 2 in correspondence to the
detection of a photon in arm 1.  This "heralded photon" was then
measured by a silicon avalanche photodiode detector (SPCM-AQR-15,
Perkin Elmer) preceded by an iris and an interference filter (IF) at
702 nm, 4 nm FWHM, inserted with the purpose of reducing the stray
light. The quantum efficiency of the detection apparatus (including
IF and iris) was measured to be $20  \%$ by using the PDC
calibration scheme (see \cite{pdccs}). Lower quantum efficiencies
were simulated by inserting calibrated neutral optical filters on
the optical path. A comparison of the observed frequencies $f_{\nu}$
with the theoretical curve ($1- \eta_{\nu}$) is presented in the
inset of Fig. \ref{f:fock}.  The photon distribution has been
reconstructed using $K=34$ different values of the quantum
efficiency from $\eta_{\nu}\simeq 0$ to $\eta_\nu\simeq 20\%$ with
$n_\nu=10^6$ runs for each $\eta_\nu$.  Results at iteration
$i=10^6$ are shown in Fig. \ref{f:fock}. As expected, the PDC
heralded  photon state largely agrees with a single photon Fock
state. However, also a small two photons component and a vacuum one
are observed.  The $\rho_2$ contribution is expected, by estimating
the probability that a second photon randomly enters the detection
window, to be $1.85 \%$ of $\rho_1$, in agreement with what
observed. A non zero $\rho_0$ is also expected due to background.
This quantity can be evaluated to correspond to $(2.7 \pm 0.2) \%$
by measuring the counts when the polarization of the pump beam is
rotated to avoid generation of parametric fluorescence. Also this
estimate is in good agreement with the reconstructed $\rho_0$.
A second example is represented by a strongly attenuated coherent
state, which has been produced by a He-Ne laser beam attenuated to
photon-counting regime  by insertion of neutral filters.  Also in
this case the same silicon avalanche photodiode detector was used.
The reconstructed distribution (with $K=15$ different values of the
quantum efficiency from $\eta_{\nu}\simeq 0$ to $\eta_\nu\simeq
66\%$ with $n_\nu=10^6$ runs for each $\eta_\nu$) agrees well with
what expected for a coherent state with average number of photons
$|\alpha|^2 \simeq 0.02$. In the inset of Fig. \ref{f:fock} the
frequencies $f_{\nu}$ as a function of $\eta_{\nu}$ are compared
with the theoretical prediction $p_\nu = \exp\{-\eta_\nu
|\alpha|^2\} \simeq 1- \eta_\nu |\alpha|^2$. Notice that in this
case we do not have IF or irises in front of the detector
and all the other attenuations can be included in the generation of
the state: thus the highest quantum efficiency is taken to be $66
\%$ as declared by the manufacturer data-sheet for the
photodetector.
\par
In the pulsed domain, we have measured three different optical
states generated starting from the third harmonics (349 nm, 4.45 ps)
of a cw mode-locked Nd:YLF laser regeneratively amplified at a
repetition rate of 500 Hz (High Q Laser). For all the measurements,
the light was delivered to a photo multiplier tube (PMT, Burle 8850)
through a multi mode fiber (100 $\mu$m core diameter). Although the
PMT has the capability of counting the number of photoelectrons
produced by one or more photons \cite{burle}, for the present
application we used it in a Geiger-like configuration, by setting a
threshold to discriminate on/off events.
%%%%%%
\begin{figure}[h]
\includegraphics[width=0.3\textwidth]{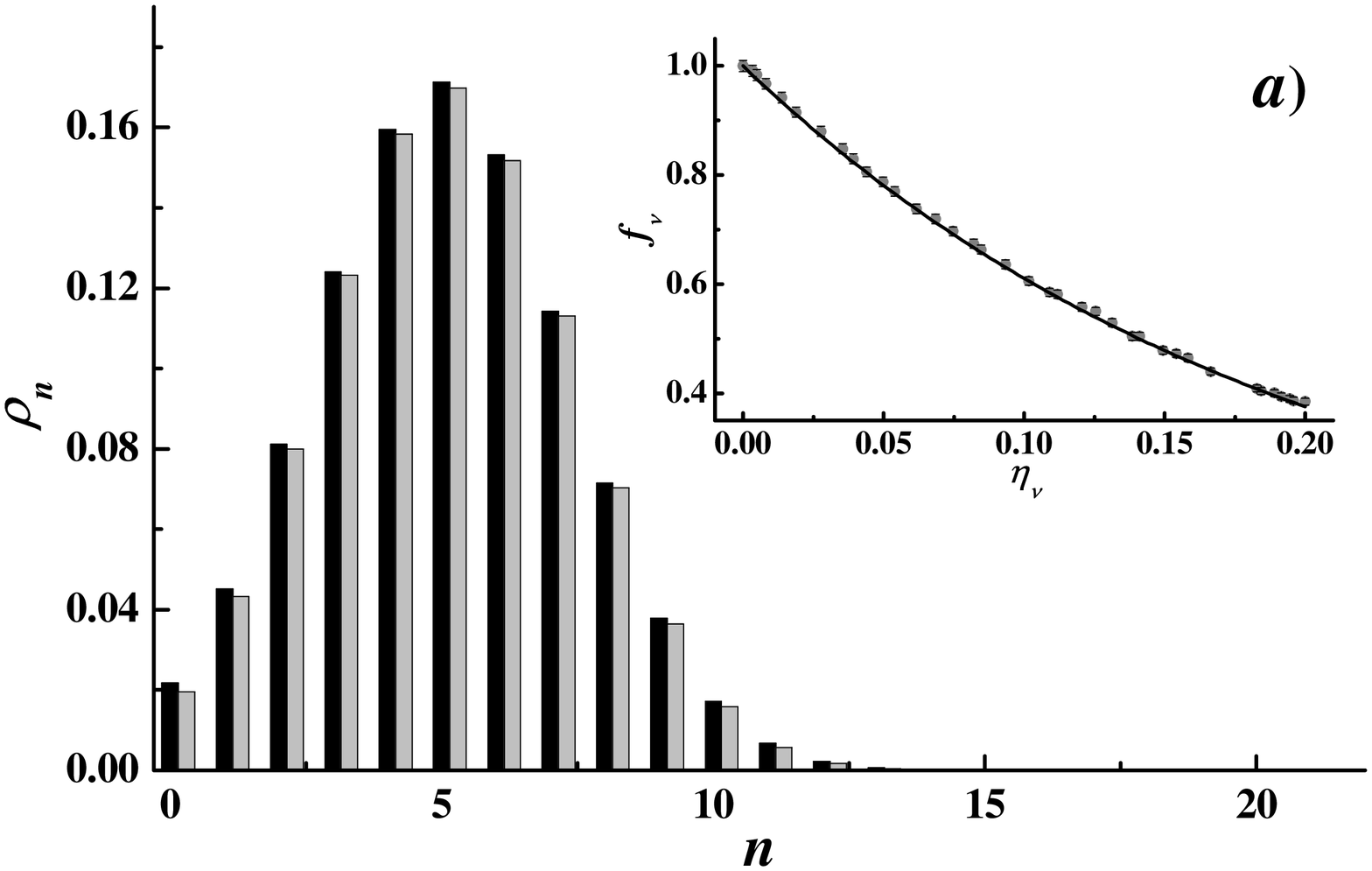}
\includegraphics[width=0.3\textwidth]{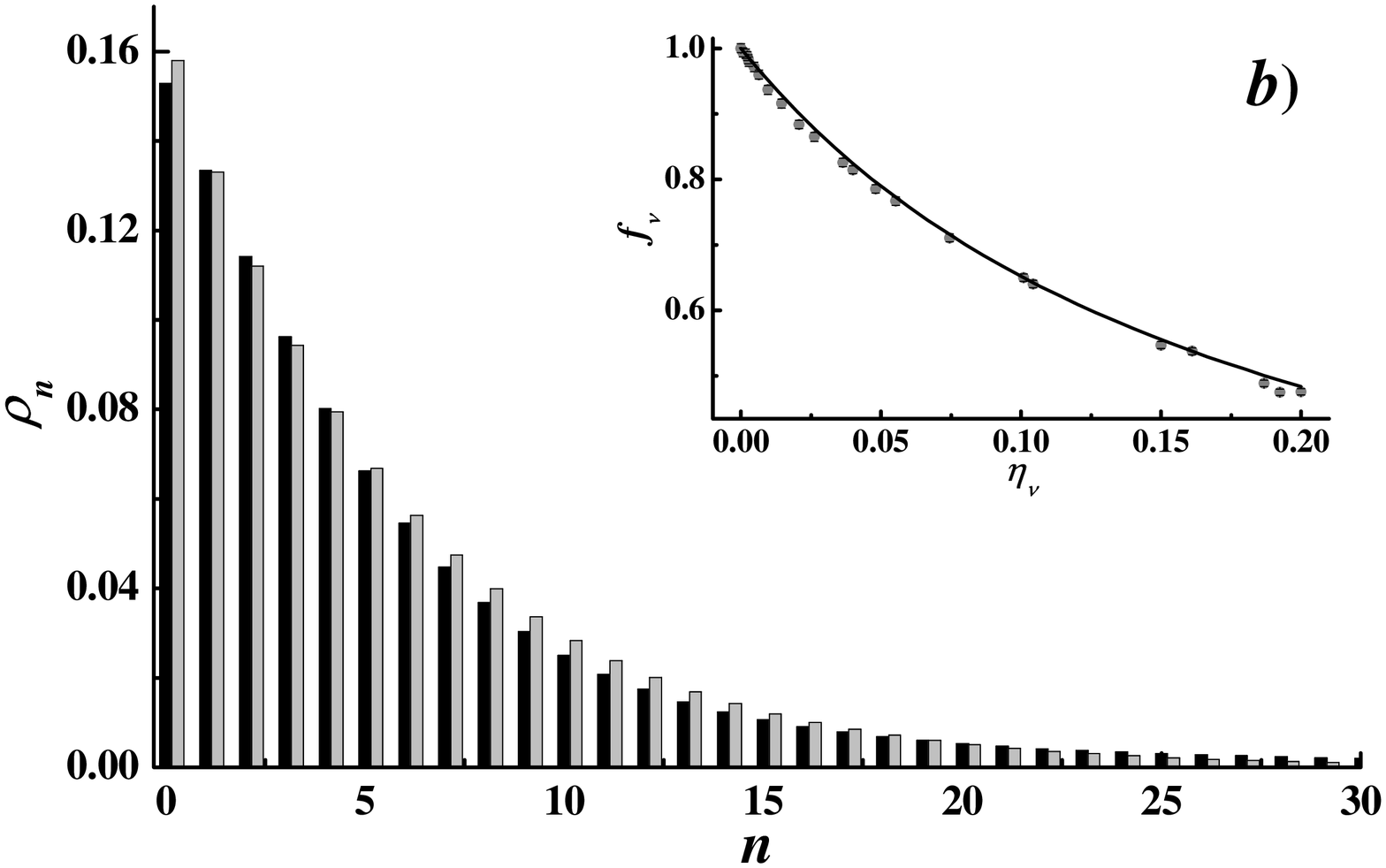}
\includegraphics[width=0.3\textwidth]{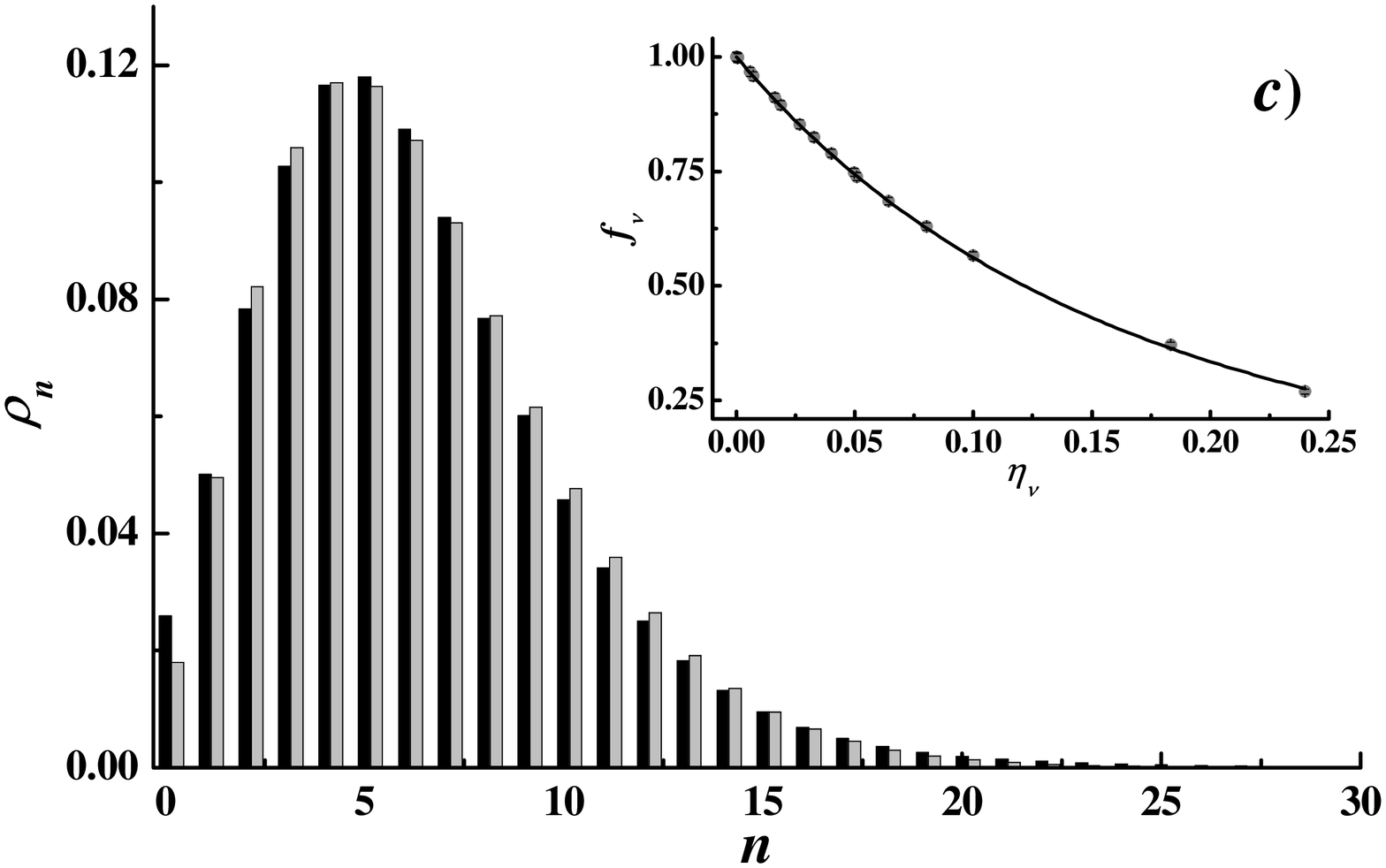}
\caption{Reconstructed photon distribution (black bars)
and best theoretical fit (grey bars) for three different
states in the pulsed regime: a) laser pulse, b) diffused
laser pulse, c) multimode state produced in type I PDC.
Insets: Experimental frequency $f_{\nu}$ data in function
of the quantum efficiency $\eta_{\nu}$ and theoretical model
for each one of the states.\label{f:total}}
\end{figure}
%%%%%%
Furthermore, by using the detector in the regime of linear response,
the knowledge of its photocathode quantum efficiency is sufficient
ti determine the $\eta_\nu$ values.
\par
The first measurement was performed on the pulse emerging from the
laser source. Due to the pulsed nature of the source, we do not
expect to recover a true Poissonian statistics. Rather, we expect
a Gaussian distribution $\varrho_{n,G}$ \cite{lasergaus} with mean
value $N$ and variance $N+\sigma^2$ which takes into account the
presence  of noise. $N$ is the photon
mean value and $\sigma^2/N$ is as a measure of the
deviation from Poissonian statistics. In Fig. \ref{f:total} a) we
show the photon distribution $\varrho_n$, reconstructed at the
$i=50000$ iteration of the ML algorithm, along with the best fit
obtained with the model $\varrho_{n,G}$ (fitting parameters $N= 4.88$
and $\sigma^2 = 0.63$). The inset of the figure compares the
experimental frequency $f_{\nu}$ data ($K=37$ values of $\eta$,
$n_\nu=10^4$ runs for each $\eta$) as a function of $\eta_\nu$ with
the theoretical values calculated through (\ref{p0}) and the
parameters given by the fit of the photon distribution. Both the
reconstructed distribution and the experimental frequencies agrees
very well with the above Gaussian model. The fidelity of the
reconstruction is $G\simeq 0.998$. Using the estimated value of
$\sigma^2$, a deviation of about $13\%$ from the Poissonian statistics
can be derived for the laser photon number distribution.
\par
A second measurement was performed on laser pulses diffused by a
moving ground glass. If the photons are collected from within an
area of spatial coherence, the system acts as a pseudo-thermal
source, whose photon number distribution is given by $\varrho_{n,T}
= N^n (N+1)^{-n-1}$. Figure \ref{f:total} b) shows the photon
distribution $\varrho_n$, as reconstructed at the $400th$ iteration
and the best fit of the data with $\varrho_{n,T}$ ($N= 5.33$); the
fidelity is given by $G\simeq 0.995$. The inset of the figure
contains the experimental frequency $f_{\nu}$ data ($K=24$ values of
$\eta$, $n_\nu=10^4$ runs for each $\eta$) and their theoretical
values as calculated from (\ref{p0}).
\par
The last measurement was performed on the blue portion (420 nm) of
the down conversion fluorescence produced by a type I BBO crystal
(10 mm depth, cut at 34 deg) pumped by the laser pulse. The pump,
incident orthogonally to the crystal face, had an intensity $\sim
60$ GW/cm$^2$. In this experimental condition we expect a coherence
time of the generated field of $\sim 1$ ps, that corresponds to
measuring a convolution of 4-5 temporal modes \cite{multit}. The
photon number distribution is expected to be a multithermal
distribution of the form
\begin{equation}
\varrho_{n,M} = \frac{(n+\mu-1)!}{n!(\mu-1)!
(1+N/\mu)^n(1+\mu/N)^{\mu}}\: , \label{multi}\:
\end{equation}
where $\mu$ is the number of temporal modes. The photon distribution
reconstructed at the $i=1500$ iteration, is shown in Fig.
\ref{f:total} c) along with the best fit of the data using
(\ref{multi}) ($N= 6.17$ and $\mu = 5$); the fidelity of the
reconstruction is given by $G\simeq 1$. In the inset of the figure
we show the experimental frequency $f_{\nu}$ data ($K=18$ values of
$\eta$, $n_\nu=10^4$ runs for each $\eta$) and their theoretical
values as calculated according to (\ref{p0}).
\par
As a comment to
the experimental results in the pulsed regime, we note that the best
reconstruction of the photon distribution is achieved at a different
number of iterations for the three different measured optical
states, and that the absolute error $\varepsilon$ does not approach
the same value. This is due to the presence of excess noise in our
measurements, since the stability and the repetition rate of our
source (500 Hz) limits to $n_\nu\sim 10^4$ the number of runs for
each value of the quantum efficiency \cite{pcount}. The choice of
the best iteration to stop the algorithm is driven by the
possibility to fit the distribution with a suitable model. We stress
that there was no \emph{a-priori} decision in choosing a Gaussian
distribution for case a) or of a multithermal distribution for case
c), but, on the contrary, we followed the \emph{a-posteriori}
observation that no other distribution could fit equally well the
reconstructed data.
\par
As to the comparison of the present technique with other schemes to
reconstruct the photon distribution we have to distinguish between
cw and pulsed regime. For cw field an alternative technique is
represented by quantum homodyne tomography (QHT), which has indeed
been applied to the reconstruction of single-photon and
single-photon added states of light \cite{bellini}. The advantage of
our technique compared to QHT is twofold. On one hand QHT requires,
for the same task, a more complex apparatus and  high-efficiency
photodetectors. On the other hand, QHT is more noisy.  In fact,
homodyne data contains the whole information about the state under
investigation (not only the photon distribution) and, in turn, this
results in a more noisy determination when only part of the
information is of interest \cite{added}. In order to obtain results
such those of Fig. \ref{f:fock} QHT requires a by far larger set of
data. In the pulsed regime, where realization of QHT is still
challenging, the direct measurement of the photon statistics can be
conveniently done  by PMT's or hybrid photodetectors
\cite{burle,NIST}. However, owing to limitations of the values of
the maximum detection efficiency, results at number of photons as
low as those in Fig.~2 cannot be obtained by PMT's and are at limit
of feasibility for hybrid photodetectors. While our method is
particularly suited to measure the photon distributions of
low-intensity field, where the above detectors are not effective, it
can be always applied to higher intensity fields.
\par
In conclusion, we experimentally implemented a reconstruction
method for the photon distribution based on on/off detection at
different quantum efficiency followed by a ML iterative
algorithm. Our experimental results demonstrate that the
technique can be applied to both cw and pulsed regimes, and for a
wide range of signal energy (from single-photon states to
mesoscopic signals), a feature that makes it preferable to other
methods till now devised for reconstructing photon number distribution.
\par\noindent
This work has been supported by MIUR (FIRB RBAU01L5AZ-002 and
RBAU014CLC-002), by INFM (PRA-CLON) and by "Regione Piemonte". MGAP
thanks Zdenek Hradil, Alessandro Ferraro and Stefano Olivares for
useful discussions.
%%%%%%%%%%%%%%%%%%%%%%%%%%%%%%%%%%%%%%%%%%%%%%%

%%%%%%%%%%%%%%

\begin{thebibliography}{99}
\bibitem{burle} G. Zambra, M. Bondani, A. S. Spinelli, A. Andreoni, Rev. Sci. Instrum.
{\bf 75}, 2762 (2004).
\bibitem{NIST} E. Hergert, Single Photon Detector Workshop, Gaithersburg, NIST (2003).
\bibitem{xxx} J. Kim, S. Takeuchi, Y. Yamamoto, and H.H. Hogue, Appl.
Phys. Lett. {\bf 74}, 902 (1999); A. Peacock, P. Verhoeve, N. Rando,
A. van Dordrecht, B. G. Taylor, C. Erd, M. A. C. Perryman, R. Venn,
J. Howlett, D. J. Goldie, J. Lumley, and M. Wallis, Nature {\bf
381}, 135 (1996).
\bibitem{serg} G. Di Giuseppe, A. V. Sergienko, B. E. A. Saleh, and
M. C. Teich in {\em Quantum Information and Computation}, E. Donkor,
A. R. Pirich, and H. E. Brandt Eds., Proceedings of the SPIE {\bf
5105}, 39 (2003).
\bibitem{mun} M. Munroe, D. Boggavarapu, M. E. Anderson, and M. G. Raymer,
Phys. Rev. A {\bf 52}, R924 (1995); Y. Zhang, K. Kasai, and M.
Watanabe, Opt. Lett. {\bf 27}, 1244 (2002).
\bibitem{raymerLNP} M. Raymer, M. Beck in {\em Quantum states estimation},
M. G .A Paris and J. \v{R}eh\'{a}\v{c}ek Eds., Lect. Not. Phys. {\bf 649}
(Springer, Berlin-Heidelberg, 2004).
\bibitem{rev} F. Zappa, A. L. Lacaita, S. D. Cova, and
P. Lovati, Opt. Eng. {\bf 35}, 938 (1996); D. Achilles, C.
Silberhorn, C. $\acute{\rm{S}}$liwa, K. Banaszek, and I. A.
Walmsley, Opt. Lett. {\bf 28}, 2387 (2003).
\bibitem{mogy} D. Mogilevtsev, Opt. Comm {\bf 156}, 307 (1998); Acta Phys.
Slov. {\bf 49}, 743 (1999).
\bibitem{pcount} A. R. Rossi, S. Olivares, and M. G. A. Paris,
Phys. Rev. A {\bf 70}, 055801 (2004).
\bibitem{ar} A. R. Rossi and M. G. A. Paris, Eur. Phys. J. D {\bf 32}, 223 (2005).
\bibitem{olom} J.  $\check{\rm R}$eh$\acute{\rm a}\check{\rm c}$ek, Z. Hradil,
O. Haderka, J.  Pe$\check{\rm r}$ina, Jr., and M.  Hamar, Phys. Rev. A {\bf 67},
061801(R) (2003); O. Haderka, M. Hamar, J. Pe$\check{\rm r}$ina, Eur. Phys. Journ.
D {\bf 28}, (2004).
\bibitem{kb}K. Banaszek, I. A. Walmsley, Opt. Lett. {\bf 28}, 52 (2003).
\bibitem{EMalg} A.P. Dempster, N.M. Laird, D.B. Rubin, J. R. Statist. Soc. B
{\bf 39}, 1 (1977); Y. Vardi and D. Lee, J. R. Statist. Soc. B
{\bf 55}, 569 (1993); R. A. Boyles, J. R. Statist. Soc. B {\bf 45},
47 (1983).
\bibitem{pdccs} G. Brida, M. Genovese, and C. Novero, J. Mod. Opt. {\bf 47}, 2099
(2000), and Refs. therein.
\bibitem{multit} F. Paleari, A. Andreoni, G. Zambra, M. Bondani.
Opt. Expr. {\bf 13}, 2816 (2004).
\bibitem{PMTexamp} G. F. Knoll, {\em Radiation detection and measurement}.
Second Edition (John Wiley \& Sons, New York, 1989).
\bibitem{lasergaus} R. Loudon, {\em The Quantum Theory of Light}, (Oxford
University Press, New York, 2000).
\bibitem{bellini} A. Zavatta, S. Viciani, M. Bellini, Science {\bf 306},
660 (2004).
\bibitem{added} G. M. D'Ariano, M. G. A. Paris, Phys. Lett. A {\bf 233} 49
(1997).
\end{thebibliography}
\end{document}